\documentclass[prl,aps,superscriptaddress,twocolumn,floatfix]{revtex4-1}
\usepackage{dcolumn} \usepackage{graphicx} \usepackage{amsmath} 
\usepackage{amssymb} \usepackage{siunitx} \usepackage{amsfonts} 
\begin{document}

\title{Resolving Dynamic Properties of Polymers through Coarse-Grained
Computational Studies} 
\author{K.\ Michael Salerno} 
\affiliation{Sandia National Laboratories, Albuquerque, NM, 87185} 
\author{Anupriya Agrawal}
\affiliation{Department of Mechanical Engineering and
Materials Science, Washington University, St. Louis, MO 63130} 
\affiliation{Department of Chemistry, Clemson University, Clemson, SC 29634} 
\author{Dvora Perahia} 
\affiliation{Department of Chemistry, Clemson University, Clemson, SC 29634} 
\author{Gary S. Grest} 
\affiliation{Sandia National Laboratories, Albuquerque, NM, 87185}

\begin{abstract}
Coupled length and time scales determine the dynamic behavior of polymers and
underlie their unique viscoelastic properties.  To resolve the long-time
dynamics it is imperative to determine which time and length scales must be
correctly modeled. Here we probe the degree of coarse graining required to
simultaneously retain significant atomistic details and access large length and
time scales. The degree of coarse graining in turn sets the minimum length
scale instrumental in defining polymer properties and dynamics.  Using linear
polyethylene as a model system, we probe how coarse graining scale affects the
measured dynamics.  Iterative Boltzmann inversion is used to derive
coarse-grained potentials with 2-6 methylene groups per coarse-grained bead
from a fully atomistic melt simulation. We show that atomistic detail
is critical to capturing large scale dynamics.  Using these models we simulate
polyethylene melts for times over 500 $\mu s$ to study the viscoelastic
properties of well-entangled polymer chains.  \end{abstract} \maketitle

Polymer properties depend on a wide range of coupled length and time scales,
with unique viscoelastic properties stemming from interactions at the atomistic
level.  The need to probe polymers across time and length scales to capture
polymer behavior makes probing dynamics, and particularly computational
modeling, inherently challenging.  With increasing molecular weight, polymer
melts become highly entangled and the long-time diffusive regime becomes
computationally inaccessible using atomistic simulations.  In these systems the
diffusive time scale increases with polymerization number $N$ faster than
$N^3$, becoming greater than $10^{10}$ times larger than the shortest time
scales even for modest molecular weight polymers.  While it is clear that the
largest lengths scales of polymer dynamics are controlled by entanglements, the
shortest time and length scales required to resolve dynamic properties are not
obvious. This knowledge is critical for developing models that can transpose
atomistic details into the long time scales needed to model long, entangled
polymer chains.  

One path to overcoming this computational challenge is to coarse grain the
polymer, reducing the number of degrees of freedom and increasing the
fundamental time scale.  The effectiveness of this process depends on retaining
the smallest length scale essential to capturing the polymer dynamics.  The
process of coarse graining amounts to combining groups of atoms into pseudoatom
beads and determining the bead interaction potentials \cite{Klein-review.2004,
Liu.2013-review}.  Simple models like the bead-spring model \cite{Kremer.1990},
capture characteristics described by scaling theories, but disregard atomistic
details and cannot quantitatively describe properties like structure, local
dynamics or densities.  Immense efforts have been made to systematically coarse
grain polymers and bridge the gap of time and length scales while retaining
atomistic characteristics \cite{MP.2002}.  One critical issue underlying the
coarse graining process is the degree to which a polymer can be coarse grained
while still appropriately capturing polymer properties and dynamics 
\cite{Padding.2011-review}.  The current study probes the effects of the degree
of coarse graining of polymers on their dynamic and static properties.

\begin{figure}
\includegraphics[width=2.40in]{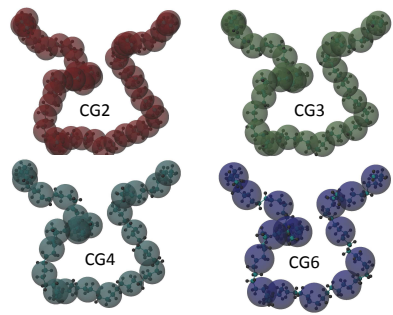}
\caption{A single C$_{96}$H$_{194}$ PE chain represented with increasing degree
of coarse graining $\lambda= 2$, 3, 4, and 6 methylene groups per CG bead. Bead
diameter corresponds to the minimum in the nonbonded interaction for each CG
model. } \label{f:cartoon} \end{figure}

With the vast efforts to coarse grain polymers, many models have emerged
with differences in the number of atoms combined in each bead and the procedure
for determining the interaction potentials.  One of the most common
coarse-grained (CG) models for polymers is the united atom (UA) model, which
combines each CH$_n$ group into a pseudoatom. The UA interaction parameters are
determined phenomenologically to reproduce physical properties such as
densities and critical temperatures \cite{Yoon.1995,NERD.1998, Martin.1998,
Mondello.1995}. Another model commonly used is the MARTINI model, which
utilizes the same approach, matching bulk densities and compressiblities of
short alkane chains at a larger scale of four CH$_2$ groups per CG bead
\cite{Marrink.2007}. More advanced methods such as force matching, iterative
Boltzmann inversion, and optimized relative entropy \cite{Izvekov.2005,
Shell.2008, Junghans.2011} have recently been developed to incorporate
atomistic detail into the CG model.  With these methods there is an open
question as to the number of atoms to represent by a single bead and the effect
of this coarse-graining scale on the measured properties of the system. One
critical physics question remains unresolved: namely defining the shortest
length scale in a polymer that is fundamental to the macroscopic dynamics and
properties.\cite{Klein-review.2004, Kremer.1990, MP.2002, Yoon.1995, NERD.1998,
Martin.1998, Mondello.1995, Marrink.2007, Izvekov.2005, Shell.2008,
Junghans.2011}  Here, this issue is addressed through the development of CG
models with increasing degree of coarse graining using iterative Boltzmann
inversion.  By examining how well these CG models describe both the static and
dynamic properties of a polymer melt, using polyethylene (PE) as a model
system, we probe this outstanding question.  The backbone of PE consists of
-CH$_2$- methylene groups that provide a natural unit or scale for coarse
graining.  Though the chemical structure of PE is simple, it is a thermoplastic
material useful in a large number of applications, with tunable mechanical
properties determined by the degree of branching.  

Polyethylene chains have previously been studied using CG models with beads of
$\lambda\ =\ 3-48$ methylene groups per bead \cite{Fukunaga.2002, Padding.2001,
Ashbaugh.2005, Guerrault.2004, Chen.2006, Aleman.2007}. These studies were able
to capture the radius of gyration as a function of molecular weight and the
pair correlation function between CG beads. As most of these studies used a
large degree of coarse graining ($\lambda \sim 20$) to study dynamical
properties, an extra constraint was needed to prevent chains cutting through
each other \cite{Padding.2002}. With this extra constraint, the mean squared
displacement (MSD), stress autocorrelation function and shear viscosity of
linear and branched PE \cite{Padding.2002,Padding.2003,Liu.2013} have been
studied for long, entangled chains. However, these studies did not account for
or study the effects of the coarse-graining degree $\lambda$ on dynamic
properties.  

Here for the first time, we elucidate the effect of coarse-graining degree on
the ability to capture both the structure and dynamics of PE.  We are able to
capture polymer chain dynamics for lengths up to C$_{1920}$H$_{3842}$ and time
scales of 400 $\mu s$ using models that accurately represent atomistic detail.
Accessing large length and time scales allows us to measure quantities like the
plateau modulus which depend on a hierarchy of length and time scales.

Coarse-grained beads shown in Fig.~\ref{f:cartoon} represent $\lambda$
methylene groups.  We study $\lambda\ =\ 2$, 3, 4 and 6 and refer to these
models as CG$\lambda$.  We find that for surprisingly small $\lambda$ the
chains cross and diffuse rapidly, indicating that CG features directly link to
macroscopic polymer motion.  With this result, we further probed the CG6 model
polymer including non-crossing constraints, and comparing with models with
unconstrained dynamics.

\begin{figure}
\includegraphics[width=3.30in]{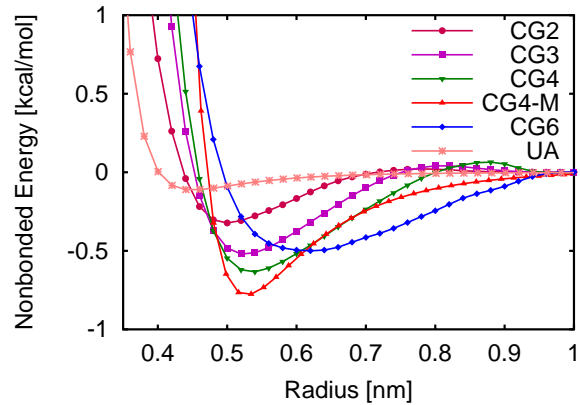}
\caption{Tabulated pair potential between CG beads.  The UA and CG4-M
potentials are included for comparison.} \label{f:potentials} \end{figure}

The tabulated CG PE potentials were derived from a single fully-atomistic
simulation of a melt of C$_{96}$H$_{194}$ PE chains at 500K.  
The study was then generalized to melts of C$_n$H$_{2n+2}$ with n=96, 480 for
the fully atomistic model, and n=96, 480, 960 and 1920 at 500 K using the CG
models.  Atomistic simulations used a version of the Optimized Potentials for
Liquid Simulations (OPLS) potential with modified dihedral coefficients that
better reproduce the properties of long alkanes \cite{Siu.2012}.  With this
modified potential the mean squared radius of gyration $\langle R_g^2 \rangle$
and end to end distance $\langle R^2 \rangle$ match experimental values
\cite{Fetters.1999, Fetters.1999.2} better than with original OPLS parameters
\cite{Jorgensen.1996}.  For the CG$\lambda$ models, $\langle R^2 \rangle $ for
n=96 chains is within 20\% of the atomistic value, while $\langle R^2 \rangle $
for the MARTINI model is 50\% too high.  Static properties for different chain
lengths are reported in the Supplement.

Tabulated CG angle and bond potentials were determined by Boltzmann inversion
of the atomistic bond and angle distributions in Fig.~S1.  Torsion terms were
omitted in all CG models, which may account for the shorter end to end
distances listed in Table SI for the CG2 model.  Tabulated nonbonded potentials
were determined by iterative Boltzmann inversion \cite{MP.2002}. The
intermolecular radial distribution function $g(r)$ from the atomistic
simulation, shown in Fig.~S2, was used as the target for iteration of the
nonbonded potentials shown in Fig.~\ref{f:potentials}.  Also shown are 6-12
Lennard Jones pair potentials for the united atom (UA) model of Yoon et al.
\cite{Yoon.1995}, and the MARTINI (CG4-M) model \cite{Marrink.2007}. The
MARTINI parameter $\epsilon$ was reduced from $0.8365$ kcal/mol to $0.803$
kcal/mol to match the density $\rho = 0.72 $ g/cm$^3$ of atomistic simulations
for $n=96$ chains.  For each CG model a pressure correction is applied to match
the density $\rho = 0.72 $ g/cm$^3$ for n=96.  The similarity in length and
energy scales between the CG4 and CG4-M models is evident in
Fig.~\ref{f:potentials}.  For each CG$\lambda$ model all beads identical
interactions, however end beads have an extra hydrogen atom mass. 

The CG6 model has a surprisingly large equilibrium bond distance relative to
the bead diameter.  Therefore a modified soft segmental repulsive potential
\cite{Sirk.2012} was added between CG beads to inhibit chain crossing.  We used
a segmental bead diameter of 0.5 nm.  This scheme increases the pressure in our
samples by about 80 atm at fixed density compared to simulations with no
constraint.  We do not re-derive the potential including the soft segmental
bead, however Fig.~S3 shows that the non-crossing bead induces only small
changes in $g(r)$ relative to a model with no constraint. By eliminating the
finest degrees of freedom, CG models allow a significantly larger time step
than atomistic models.  We use a time step $\delta$t = 20 fs for the CG6, CG4
and CG4-M models, 10 fs for the CG3 model and 2 fs for the CG2 model, compared
to 1 fs for the atomistic model.

\begin{figure}
\includegraphics[width=3.20in]{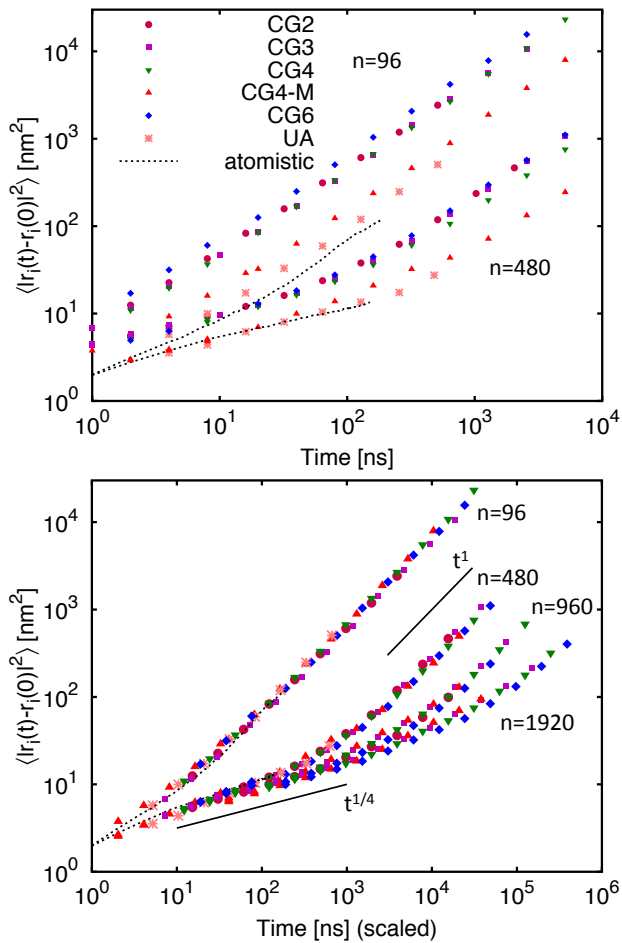}
\caption{ (a) The MSD of the inner 24 -CH$_2$- groups of each polymer chain at
500 K.  (b) Same data as in (a), scaled by $\alpha.$ The solid lines represent
the scaling predictions $t^1$ for the diffusive regime and $t^{1/4}$ for the
reptation regime.} \label{f:g1} \end{figure}

Coarse graining reduces the number of degrees of freedom in a system, creating
a smoother free-energy landscape compared with fully-atomistic simulations.
This speedup can be addressed by including frictional and stochastic forces
\cite{Padding.2001}.  This approach allows large CG scales, however rigorously
correct dissipation requires a sophisticated generalized Langevin kernel, and
simplifications are usually employed \cite{Padding.2011-review}.  It has been
shown that for small $\lambda$ CG models without added friction or stochastic
forces can be employed, but CG dynamics are significantly faster than in
atomistic simulations \cite{Maranas.2011, Maranas.2005,
Lyubimov.2013,Lyubimov.2010, Harmandaris.2014,Harmandaris.2009,Fritz.2011,
Hess.2006}.  To determine the dynamic scaling factor of the CG models we
compare the MSD of the inner 24 methylene groups (4, 6, 8 or 12 beads) for CG
models and the inner 24 carbon atoms for atomistic simulations for n=96 and 480
as shown in Fig.~\ref{f:g1} (a). The mobility of the chains in the CG models is
larger than in atomistic simulations.  By scaling the time for each of the CG
models we create a single collapsed curve for each chain length for both the
atomistic and CG data as shown in Fig.\ref{f:g1} (b).  Notably, a single
scaling factor $\alpha$ is required to collapse atomistic and CG data for each
model, independent of chain length.  As seen in Fig.~3b the MSD has reached the
diffusive regime where MSD $\sim t^1$ even for the longest chain length
$n=1920$. Over intermediate time scales, the chains show the expected $t^{1/4}$
scaling predicted by reptation theory \cite{deGennes.1980}.  These results
demonstrate that one can capture long time and length scales with CG models
while accounting for atomistic details. 

\begin{figure}
\includegraphics[width=2.50in,angle=90]{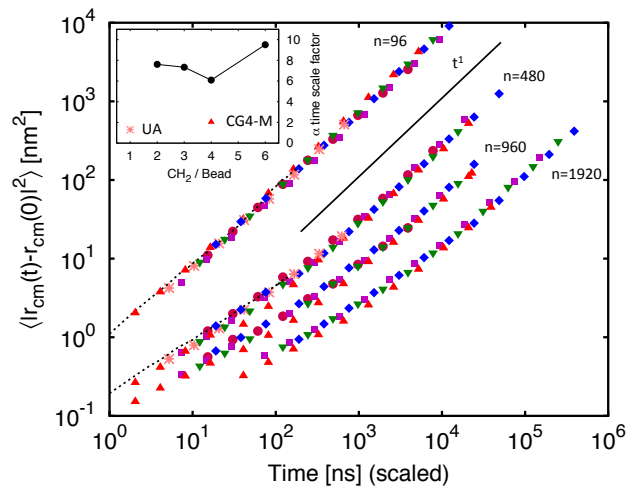}
\caption{ MSD of center of mass scaled with the same $\alpha$ scale factor as
the for the inner -CH$_2$- groups in Fig.~\ref{f:g1} (b). The solid line
has slope $t^1$.  Inset: The alpha scale factor for different coarse-grained
models and the UA and Martini models.} \label{f:g3} \end{figure}

The MSD of the center of mass was then measured to test the scaling factor
$\alpha.$  Figure \ref{f:g3} shows the MSD of the chain center of mass for
chain lengths n=96, 480, 960, and 1920.  These data have been scaled by the
same $\alpha$ as the monomer MSD, producing an excellent collapse.  The scale
factor $\alpha$ as a function of CG model is shown inset in Fig.~\ref{f:g3}
along with $\alpha$ for the MARTINI \cite{Marrink.2007} and UA \cite{Yoon.1995}
models.  Our CG potentials have a much larger time scaling factor than the
MARTINI and UA models, similar to the time-scaling factor found previously for
PE for a single $\lambda$ \cite{Maranas.2005}. Values of $\alpha$ are also
comparable to those found previously for polystyrene, modeled at a similar
coarse-graining level \cite{Harmandaris.2009}.  The UA model has long been
considered approximate to the fully-atomistic simulation and indeed is
$\approx$40\% faster than fully-atomistic simulations.  Interestingly, the
time scaling factor is not monotonic in CG level, with the CG2 and CG6 models
exhibiting the largest speedup.  The potential depths in
Fig.~\ref{f:potentials}, relate to the value of $\alpha,$ as described
previously by Depa and Maranas \cite{Maranas.2005}.

The polymer entanglement mass M$_e$ governs many properties of the polymer melt
and provides information about chain mobility within the polymer mesh.
Experimentally, M$_e = \rho RT/G^0_N$ is determined from the plateau modulus
$G_N^0$ of the stress relaxation function $G(t)$ \cite{Fetters.1999}.
Experimental values for polyethylene are 1.6-2.5 MPa, corresponding to $M_e$ of
1300-2000 g/mol \cite{Vega.2004, Raju.1979, Fetters.1999, Fetters.1999.2}.

\begin{figure}
\includegraphics[width=3.60in]{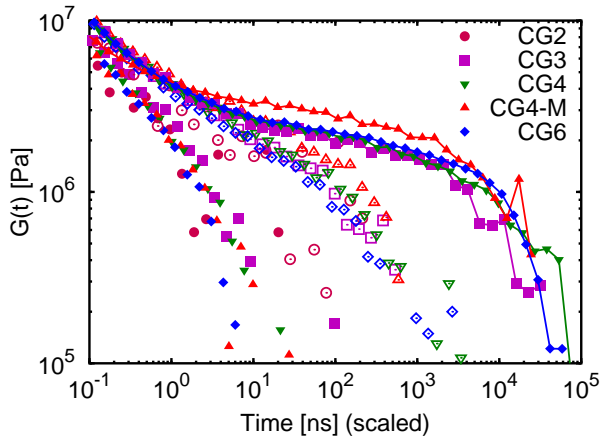}
\caption{ Modulus $G(t) $ for each of the CG polymer models at 500K.  Filled
and open symbols represent the n=96 and n=480 chain length, respectively.
Solid lines represent the n=1920 length, while n=960 chains are omitted for
clarity.  } \label{f:plateau} \end{figure}

The relaxation modulus in each of our CG models was measured for the four
different chain lengths via equilibrium stress correlations using the
Green-Kubo relation $G(t) = (V/k_BT) \langle
\sigma_{\alpha\beta}(t)\sigma_{\alpha\beta}(0) \rangle$ where
$\sigma_{\alpha\beta}$ are the off-diagonal components $xy$, $xz$, and $yz$ of
stress.  Figure \ref{f:plateau} shows $G^0_N$ for each of
the CG models for n=96 and 480 and for $\lambda \ge 3$ for n = 1920.  The times
for each model have been scaled by the corresponding value of $\alpha.$  Though
it shows similar behavior, the UA model is omitted because the zero-pressure
density is higher than the other models, making comparison difficult.  The
curves collapse for the short-time $t^{-1/2}$ regime, with longer, more
entangled chains forming progressively more distinct plateau regions.  The
plateau modulus is measured as the value of the relaxation modulus in the
plateau region, roughly between 20 and 600 ns.  Using the longest chain length,
n = 1920, the plateau modulus $G^0_N = 2.2 \pm 0.3$ MPa for CG6, $2.1 \pm 0.3 $
MPa for CG4 and $2.1 \pm 0.7 $ MPa for CG3, all within the experimental range.
For CG4-M $G^0_N$ is significantly higher $G^0_N = 3.7 \pm 0.4 $ MPa and does
not agree with experimental values.  Uncertainties are measured by dividing the
datasets in two and measuring the variation in the plateau value.  Similar
plateau modulus values were found by Padding and Briels \cite{Padding.2002} for
$n \le 1000$ with $\lambda=20$ model with a non-crossing constraint.    

\begin{figure}
\includegraphics[width=3.60in]{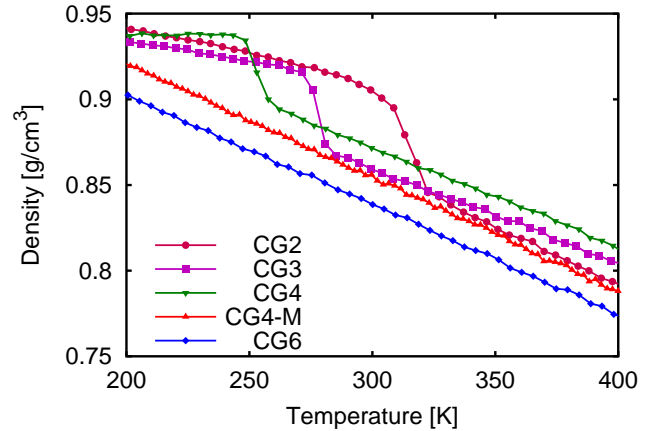}
\caption{Density versus temperature for the n=96 samples cooled at 1.4K/ns.}
\label{f:npt} \end{figure}

The thermal expansion coefficient is another way to assess the validity of the
CG models.  The linear thermal expansion coefficients for the n=96 samples in
the temperature range from 495K to 480K are 5.0, 3.4, 3.5, 3.3, 3.4 $\times
10^{-4} T^{-1}$ respectively for the CG2, CG3, CG4, CG4-M, and CG6 models,
compared with $3.1 \times 10^{-4} T^{-1} $ for the atomistic model.  Hence, all
the CG models with $\lambda\ge 3$ agree with the atomistic thermal expansion. 

Although one can derive CG potentials using multiple temperatures
\cite{Moore.2014}, our CG models are developed in the traditional way at a
single temperature.  Hence there is uncertainty about the validity of these
models away from the chosen point \cite{Carbone.2008}.  Shown in
Fig.~\ref{f:npt} are temperature-density data for the CG models from 400K to
200K.  For $\lambda \le 4$, the curves show a pronounced density increase
between 250K and 330K, corresponding to a semi-crystalline state.  The CG6
model does not capture crystallinity and we expect that coarser models will
not, either.  Previous studies of bead-spring polymer models indicated that a
commensurate bond length and bead diameter leads to crystallization
\cite{Hoy.2013}.  The melting temperature for n=96 is about 400K, so
crystallization occurs at a lower temperature than expected, yet observation of
any semi-crystalline phase is remarkable.  This surprising feature indicates
that although our potentials are derived at 500K they may be useful away from
this temperature.

Here we have shown that the smallest length scale needed in the hierarchy of
length scales to correctly describe macroscopic behavior and properties is
rather small, between 4 and 6 monomers. The CG4 model offers a speedup of more
than four orders of magnitude over atomistic simulations, which includes
contributions from the time step $\delta t,$ scale factor $\alpha,$ and
$(3\lambda)^2$ reduction in the number of pairwise interactions. The speedup of
the CG4 model is about three times faster than the CG3 model and is comparable
the CG6 model.  However because the CG4 model does not require a crossing
constraint, we prefer this model.  With this realized speedup polymers as long
as $n = 1920$  can now be simulated for over 500 $\mu s.$ Reaching this time
scale allows probing some of the most unique and intrinsic properties of
polymers including the plateau modulus and intermediate t$^{1/4}$ scaling in
the mean squared displacement.  From the computational viewpoint, the CG models
developed significantly reduce the resources needed to study polymers for long
times.  Our results for the plateau modulus and diffusion dynamics show that
without adding extra constraints the CG4 model captures the atomistic detail
needed for correct dynamics from monomer to polymer scale.

AA and DP acknowledge financial support from Grant No.~DE-SC007908 and an
allotment of time on the Clemson University Palmetto cluster.  This research
used resources at the National Energy Research Scientific Computing Center,
which is supported by the Office of Science of the U.S.~Department of
Energy under Contract DE-AC02-05CH11231.  This work was supported by the
Sandia Laboratory Directed Research and Development Program.  Research was
carried out in part, at the Center for Integrated Nanotechnologies, a U.S.~
Department of Energy, Office of Basic Energy Sciences user facility. Sandia
National Laboratories is a multi-program laboratory managed and operated by
Sandia Corporation, a wholly owned subsidiary of Lockheed Martin Corporation,
for the U.S.~Department of Energy's National Nuclear Security Administration
under contract DE-AC04-94AL85000.
 
\bibliographystyle{apsrev4-1}

\begin{thebibliography}{45}%
\makeatletter
\providecommand \@ifxundefined [1]{%
 \@ifx{#1\undefined}
}%
\providecommand \@ifnum [1]{%
 \ifnum #1\expandafter \@firstoftwo
 \else \expandafter \@secondoftwo
 \fi
}%
\providecommand \@ifx [1]{%
 \ifx #1\expandafter \@firstoftwo
 \else \expandafter \@secondoftwo
 \fi
}%
\providecommand \natexlab [1]{#1}%
\providecommand \enquote  [1]{``#1''}%
\providecommand \bibnamefont  [1]{#1}%
\providecommand \bibfnamefont [1]{#1}%
\providecommand \citenamefont [1]{#1}%
\providecommand \href@noop [0]{\@secondoftwo}%
\providecommand \href [0]{\begingroup \@sanitize@url \@href}%
\providecommand \@href[1]{\@@startlink{#1}\@@href}%
\providecommand \@@href[1]{\endgroup#1\@@endlink}%
\providecommand \@sanitize@url [0]{\catcode `\\12\catcode `\$12\catcode
  `\&12\catcode `\#12\catcode `\^12\catcode `\_12\catcode `\%12\relax}%
\providecommand \@@startlink[1]{}%
\providecommand \@@endlink[0]{}%
\providecommand \url  [0]{\begingroup\@sanitize@url \@url }%
\providecommand \@url [1]{\endgroup\@href {#1}{\urlprefix }}%
\providecommand \urlprefix  [0]{URL }%
\providecommand \Eprint [0]{\href }%
\providecommand \doibase [0]{http://dx.doi.org/}%
\providecommand \selectlanguage [0]{\@gobble}%
\providecommand \bibinfo  [0]{\@secondoftwo}%
\providecommand \bibfield  [0]{\@secondoftwo}%
\providecommand \translation [1]{[#1]}%
\providecommand \BibitemOpen [0]{}%
\providecommand \bibitemStop [0]{}%
\providecommand \bibitemNoStop [0]{.\EOS\space}%
\providecommand \EOS [0]{\spacefactor3000\relax}%
\providecommand \BibitemShut  [1]{\csname bibitem#1\endcsname}%
\let\auto@bib@innerbib\@empty
%</preamble>
\bibitem [{\citenamefont {Nielsen}\ \emph {et~al.}(2004)\citenamefont
  {Nielsen}, \citenamefont {Lopez}, \citenamefont {Srinivas},\ and\
  \citenamefont {Klein}}]{Klein-review.2004}%
  \BibitemOpen
  \bibfield  {author} {\bibinfo {author} {\bibfnamefont {S.~O.}\ \bibnamefont
  {Nielsen}}, \bibinfo {author} {\bibfnamefont {C.~F.}\ \bibnamefont {Lopez}},
  \bibinfo {author} {\bibfnamefont {G.}~\bibnamefont {Srinivas}}, \ and\
  \bibinfo {author} {\bibfnamefont {M.~L.}\ \bibnamefont {Klein}},\ }\href
  {http://stacks.iop.org/0953-8984/16/i=15/a=R03} {\bibfield  {journal}
  {\bibinfo  {journal} {J. Phys. Condens. Mat.}\ }\textbf {\bibinfo {volume}
  {16}},\ \bibinfo {pages} {R481} (\bibinfo {year} {2004})}\BibitemShut
  {NoStop}%
\bibitem [{\citenamefont {Li}\ \emph {et~al.}(2013)\citenamefont {Li},
  \citenamefont {Abberton}, \citenamefont {Kr{\"o}ger},\ and\ \citenamefont
  {Liu}}]{Liu.2013-review}%
  \BibitemOpen
  \bibfield  {author} {\bibinfo {author} {\bibfnamefont {Y.}~\bibnamefont
  {Li}}, \bibinfo {author} {\bibfnamefont {B.~C.}\ \bibnamefont {Abberton}},
  \bibinfo {author} {\bibfnamefont {M.}~\bibnamefont {Kr{\"o}ger}}, \ and\
  \bibinfo {author} {\bibfnamefont {W.~K.}\ \bibnamefont {Liu}},\ }\href@noop
  {} {\bibfield  {journal} {\bibinfo  {journal} {Polymers}\ }\textbf {\bibinfo
  {volume} {5}},\ \bibinfo {pages} {751} (\bibinfo {year} {2013})}\BibitemShut
  {NoStop}%
\bibitem [{\citenamefont {Kremer}\ and\ \citenamefont
  {Grest}(1990)}]{Kremer.1990}%
  \BibitemOpen
  \bibfield  {author} {\bibinfo {author} {\bibfnamefont {K.}~\bibnamefont
  {Kremer}}\ and\ \bibinfo {author} {\bibfnamefont {G.~S.}\ \bibnamefont
  {Grest}},\ }\href@noop {} {\bibfield  {journal} {\bibinfo  {journal} {J.
  Chem. Phys.}\ }\textbf {\bibinfo {volume} {92}},\ \bibinfo {pages} {5057}
  (\bibinfo {year} {1990})}\BibitemShut {NoStop}%
\bibitem [{\citenamefont {M\"{u}ller-Plathe}(2002)}]{MP.2002}%
  \BibitemOpen
  \bibfield  {author} {\bibinfo {author} {\bibfnamefont {F.}~\bibnamefont
  {M\"{u}ller-Plathe}},\ }\href
  {http://dx.doi.org/10.1002/1439-7641(20020916)3:9<754::AID-CPHC754>3.0.CO;2-U}
  {\bibfield  {journal} {\bibinfo  {journal} {Chem. Phys. Chem.}\ }\textbf
  {\bibinfo {volume} {3}},\ \bibinfo {pages} {754} (\bibinfo {year}
  {2002})}\BibitemShut {NoStop}%
\bibitem [{\citenamefont {Padding}\ and\ \citenamefont
  {Briels}(2011)}]{Padding.2011-review}%
  \BibitemOpen
  \bibfield  {author} {\bibinfo {author} {\bibfnamefont {J.~T.}\ \bibnamefont
  {Padding}}\ and\ \bibinfo {author} {\bibfnamefont {W.~J.}\ \bibnamefont
  {Briels}},\ }\href {http://stacks.iop.org/0953-8984/23/i=23/a=233101}
  {\bibfield  {journal} {\bibinfo  {journal} {Journal of Physics: Condensed
  Matter}\ }\textbf {\bibinfo {volume} {23}},\ \bibinfo {pages} {233101}
  (\bibinfo {year} {2011})}\BibitemShut {NoStop}%
\bibitem [{\citenamefont {Paul}\ \emph {et~al.}(1995)\citenamefont {Paul},
  \citenamefont {Yoon},\ and\ \citenamefont {Smith}}]{Yoon.1995}%
  \BibitemOpen
  \bibfield  {author} {\bibinfo {author} {\bibfnamefont {W.}~\bibnamefont
  {Paul}}, \bibinfo {author} {\bibfnamefont {D.~Y.}\ \bibnamefont {Yoon}}, \
  and\ \bibinfo {author} {\bibfnamefont {G.~D.}\ \bibnamefont {Smith}},\ }\href
  {\doibase http://dx.doi.org/10.1063/1.469740} {\bibfield  {journal} {\bibinfo
   {journal} {J. Chem. Phys.}\ }\textbf {\bibinfo {volume} {103}},\ \bibinfo
  {pages} {1702} (\bibinfo {year} {1995})}\BibitemShut {NoStop}%
\bibitem [{\citenamefont {Nath}\ \emph {et~al.}(1998)\citenamefont {Nath},
  \citenamefont {Escobedo},\ and\ \citenamefont {de~Pablo}}]{NERD.1998}%
  \BibitemOpen
  \bibfield  {author} {\bibinfo {author} {\bibfnamefont {S.~K.}\ \bibnamefont
  {Nath}}, \bibinfo {author} {\bibfnamefont {F.~A.}\ \bibnamefont {Escobedo}},
  \ and\ \bibinfo {author} {\bibfnamefont {J.~J.}\ \bibnamefont {de~Pablo}},\
  }\href {\doibase http://dx.doi.org/10.1063/1.476429} {\bibfield  {journal}
  {\bibinfo  {journal} {J. Chem. Phys.}\ }\textbf {\bibinfo {volume} {108}},\
  \bibinfo {pages} {9905} (\bibinfo {year} {1998})}\BibitemShut {NoStop}%
\bibitem [{\citenamefont {Martin}\ and\ \citenamefont
  {Siepmann}(1998)}]{Martin.1998}%
  \BibitemOpen
  \bibfield  {author} {\bibinfo {author} {\bibfnamefont {M.~G.}\ \bibnamefont
  {Martin}}\ and\ \bibinfo {author} {\bibfnamefont {J.~I.}\ \bibnamefont
  {Siepmann}},\ }\href@noop {} {\bibfield  {journal} {\bibinfo  {journal} {J.
  Phys. Chem. B}\ }\textbf {\bibinfo {volume} {102}},\ \bibinfo {pages} {2569}
  (\bibinfo {year} {1998})}\BibitemShut {NoStop}%
\bibitem [{\citenamefont {Mondello}\ and\ \citenamefont
  {Grest}(1995)}]{Mondello.1995}%
  \BibitemOpen
  \bibfield  {author} {\bibinfo {author} {\bibfnamefont {M.}~\bibnamefont
  {Mondello}}\ and\ \bibinfo {author} {\bibfnamefont {G.~S.}\ \bibnamefont
  {Grest}},\ }\href {\doibase http://dx.doi.org/10.1063/1.470344} {\bibfield
  {journal} {\bibinfo  {journal} {J. Chem. Phys.}\ }\textbf {\bibinfo {volume}
  {103}},\ \bibinfo {pages} {7156} (\bibinfo {year} {1995})}\BibitemShut
  {NoStop}%
\bibitem [{\citenamefont {Marrink}\ \emph {et~al.}(2007)\citenamefont
  {Marrink}, \citenamefont {Risselada}, \citenamefont {Yefimov}, \citenamefont
  {Tieleman},\ and\ \citenamefont {de~Vries}}]{Marrink.2007}%
  \BibitemOpen
  \bibfield  {author} {\bibinfo {author} {\bibfnamefont {S.~J.}\ \bibnamefont
  {Marrink}}, \bibinfo {author} {\bibfnamefont {H.~J.}\ \bibnamefont
  {Risselada}}, \bibinfo {author} {\bibfnamefont {S.}~\bibnamefont {Yefimov}},
  \bibinfo {author} {\bibfnamefont {D.~P.}\ \bibnamefont {Tieleman}}, \ and\
  \bibinfo {author} {\bibfnamefont {A.~H.}\ \bibnamefont {de~Vries}},\
  }\href@noop {} {\bibfield  {journal} {\bibinfo  {journal} {J. Phys. Chem. B}\
  }\textbf {\bibinfo {volume} {111}},\ \bibinfo {pages} {7812} (\bibinfo {year}
  {2007})}\BibitemShut {NoStop}%
\bibitem [{\citenamefont {Izvekov}\ and\ \citenamefont
  {Voth}(2005)}]{Izvekov.2005}%
  \BibitemOpen
  \bibfield  {author} {\bibinfo {author} {\bibfnamefont {S.}~\bibnamefont
  {Izvekov}}\ and\ \bibinfo {author} {\bibfnamefont {G.~A.}\ \bibnamefont
  {Voth}},\ }\href@noop {} {\bibfield  {journal} {\bibinfo  {journal} {J. Phys.
  Chem. B}\ }\textbf {\bibinfo {volume} {109}},\ \bibinfo {pages} {2469}
  (\bibinfo {year} {2005})}\BibitemShut {NoStop}%
\bibitem [{\citenamefont {Shell}(2008)}]{Shell.2008}%
  \BibitemOpen
  \bibfield  {author} {\bibinfo {author} {\bibfnamefont {M.~S.}\ \bibnamefont
  {Shell}},\ }\href@noop {} {\bibfield  {journal} {\bibinfo  {journal} {J.
  Chem. Phys.}\ }\textbf {\bibinfo {volume} {129}},\ \bibinfo {eid} {144108}
  (\bibinfo {year} {2008})}\BibitemShut {NoStop}%
\bibitem [{\citenamefont {R{\"u}hle}\ and\ \citenamefont
  {Junghans}(2011)}]{Junghans.2011}%
  \BibitemOpen
  \bibfield  {author} {\bibinfo {author} {\bibfnamefont {V.}~\bibnamefont
  {R{\"u}hle}}\ and\ \bibinfo {author} {\bibfnamefont {C.}~\bibnamefont
  {Junghans}},\ }\href@noop {} {\bibfield  {journal} {\bibinfo  {journal}
  {Macromol. Theor. Simul.}\ }\textbf {\bibinfo {volume} {20}},\ \bibinfo
  {pages} {472} (\bibinfo {year} {2011})}\BibitemShut {NoStop}%
\bibitem [{\citenamefont {Fukunaga}\ \emph {et~al.}(2002)\citenamefont
  {Fukunaga}, \citenamefont {Takimoto},\ and\ \citenamefont
  {Doi}}]{Fukunaga.2002}%
  \BibitemOpen
  \bibfield  {author} {\bibinfo {author} {\bibfnamefont {H.}~\bibnamefont
  {Fukunaga}}, \bibinfo {author} {\bibfnamefont {J.-i.}\ \bibnamefont
  {Takimoto}}, \ and\ \bibinfo {author} {\bibfnamefont {M.}~\bibnamefont
  {Doi}},\ }\href {\doibase http://dx.doi.org/10.1063/1.1469609} {\bibfield
  {journal} {\bibinfo  {journal} {J. Chem. Phys.}\ }\textbf {\bibinfo {volume}
  {116}},\ \bibinfo {pages} {8183} (\bibinfo {year} {2002})}\BibitemShut
  {NoStop}%
\bibitem [{\citenamefont {Padding}\ and\ \citenamefont
  {Briels}(2001)}]{Padding.2001}%
  \BibitemOpen
  \bibfield  {author} {\bibinfo {author} {\bibfnamefont {J.~T.}\ \bibnamefont
  {Padding}}\ and\ \bibinfo {author} {\bibfnamefont {W.~J.}\ \bibnamefont
  {Briels}},\ }\href@noop {} {\bibfield  {journal} {\bibinfo  {journal} {J.
  Chem. Phys.}\ }\textbf {\bibinfo {volume} {115}},\ \bibinfo {pages} {2846}
  (\bibinfo {year} {2001})}\BibitemShut {NoStop}%
\bibitem [{\citenamefont {Ashbaugh}\ \emph {et~al.}(2005)\citenamefont
  {Ashbaugh}, \citenamefont {Patel}, \citenamefont {Kumar},\ and\ \citenamefont
  {Garde}}]{Ashbaugh.2005}%
  \BibitemOpen
  \bibfield  {author} {\bibinfo {author} {\bibfnamefont {H.~S.}\ \bibnamefont
  {Ashbaugh}}, \bibinfo {author} {\bibfnamefont {H.~A.}\ \bibnamefont {Patel}},
  \bibinfo {author} {\bibfnamefont {S.~K.}\ \bibnamefont {Kumar}}, \ and\
  \bibinfo {author} {\bibfnamefont {S.}~\bibnamefont {Garde}},\ }\href
  {\doibase http://dx.doi.org/10.1063/1.1861455} {\bibfield  {journal}
  {\bibinfo  {journal} {J. Chem. Phys.}\ }\textbf {\bibinfo {volume} {122}},\
  \bibinfo {eid} {104908} (\bibinfo {year} {2005})}\BibitemShut {NoStop}%
\bibitem [{\citenamefont {Guerrault}\ \emph {et~al.}(2004)\citenamefont
  {Guerrault}, \citenamefont {Rousseau},\ and\ \citenamefont
  {Farago}}]{Guerrault.2004}%
  \BibitemOpen
  \bibfield  {author} {\bibinfo {author} {\bibfnamefont {X.}~\bibnamefont
  {Guerrault}}, \bibinfo {author} {\bibfnamefont {B.}~\bibnamefont {Rousseau}},
  \ and\ \bibinfo {author} {\bibfnamefont {J.}~\bibnamefont {Farago}},\ }\href
  {\doibase http://dx.doi.org/10.1063/1.1786917} {\bibfield  {journal}
  {\bibinfo  {journal} {J. Chem. Phys.}\ }\textbf {\bibinfo {volume} {121}},\
  \bibinfo {pages} {6538} (\bibinfo {year} {2004})}\BibitemShut {NoStop}%
\bibitem [{\citenamefont {Chen}\ \emph {et~al.}(2006)\citenamefont {Chen},
  \citenamefont {Qian}, \citenamefont {Lu}, \citenamefont {Li},\ and\
  \citenamefont {Sun}}]{Chen.2006}%
  \BibitemOpen
  \bibfield  {author} {\bibinfo {author} {\bibfnamefont {L.-J.}\ \bibnamefont
  {Chen}}, \bibinfo {author} {\bibfnamefont {H.-J.}\ \bibnamefont {Qian}},
  \bibinfo {author} {\bibfnamefont {Z.-Y.}\ \bibnamefont {Lu}}, \bibinfo
  {author} {\bibfnamefont {Z.-S.}\ \bibnamefont {Li}}, \ and\ \bibinfo {author}
  {\bibfnamefont {C.-C.}\ \bibnamefont {Sun}},\ }\href@noop {} {\bibfield
  {journal} {\bibinfo  {journal} {J. Phys. Chem. B}\ }\textbf {\bibinfo
  {volume} {110}},\ \bibinfo {pages} {24093} (\bibinfo {year}
  {2006})}\BibitemShut {NoStop}%
\bibitem [{\citenamefont {Curc\'{o}}\ and\ \citenamefont
  {Alem\'{a}n}(2007)}]{Aleman.2007}%
  \BibitemOpen
  \bibfield  {author} {\bibinfo {author} {\bibfnamefont {D.}~\bibnamefont
  {Curc\'{o}}}\ and\ \bibinfo {author} {\bibfnamefont {C.}~\bibnamefont
  {Alem\'{a}n}},\ }\href {\doibase
  http://dx.doi.org/10.1016/j.cplett.2007.01.031} {\bibfield  {journal}
  {\bibinfo  {journal} {Chem. Phys. Lett.}\ }\textbf {\bibinfo {volume}
  {436}},\ \bibinfo {pages} {189 } (\bibinfo {year} {2007})}\BibitemShut
  {NoStop}%
\bibitem [{\citenamefont {Padding}\ and\ \citenamefont
  {Briels}(2002)}]{Padding.2002}%
  \BibitemOpen
  \bibfield  {author} {\bibinfo {author} {\bibfnamefont {J.~T.}\ \bibnamefont
  {Padding}}\ and\ \bibinfo {author} {\bibfnamefont {W.~J.}\ \bibnamefont
  {Briels}},\ }\href@noop {} {\bibfield  {journal} {\bibinfo  {journal} {J.
  Chem. Phys.}\ }\textbf {\bibinfo {volume} {117}},\ \bibinfo {pages} {925}
  (\bibinfo {year} {2002})}\BibitemShut {NoStop}%
\bibitem [{\citenamefont {Padding}\ and\ \citenamefont
  {Briels}(2003)}]{Padding.2003}%
  \BibitemOpen
  \bibfield  {author} {\bibinfo {author} {\bibfnamefont {J.~T.}\ \bibnamefont
  {Padding}}\ and\ \bibinfo {author} {\bibfnamefont {W.~J.}\ \bibnamefont
  {Briels}},\ }\href@noop {} {\bibfield  {journal} {\bibinfo  {journal} {J.
  Chem. Phys.}\ }\textbf {\bibinfo {volume} {118}},\ \bibinfo {pages} {10276}
  (\bibinfo {year} {2003})}\BibitemShut {NoStop}%
\bibitem [{\citenamefont {Liu}\ \emph {et~al.}(2013)\citenamefont {Liu},
  \citenamefont {Padding}, \citenamefont {den Otter},\ and\ \citenamefont
  {Briels}}]{Liu.2013}%
  \BibitemOpen
  \bibfield  {author} {\bibinfo {author} {\bibfnamefont {L.}~\bibnamefont
  {Liu}}, \bibinfo {author} {\bibfnamefont {J.~T.}\ \bibnamefont {Padding}},
  \bibinfo {author} {\bibfnamefont {W.~K.}\ \bibnamefont {den Otter}}, \ and\
  \bibinfo {author} {\bibfnamefont {W.~J.}\ \bibnamefont {Briels}},\ }\href
  {\doibase http://dx.doi.org/10.1063/1.4811675} {\bibfield  {journal}
  {\bibinfo  {journal} {J. Chem. Phys.}\ }\textbf {\bibinfo {volume} {138}},\
  \bibinfo {eid} {244912} (\bibinfo {year} {2013})}\BibitemShut {NoStop}%
\bibitem [{Note1()}]{Note1}%
  \BibitemOpen
  \bibinfo {note} {See Supplemental Material [url], which includes Refs. \cite
  {Tuckerman.1992, Isele-Holder.2012, Plimpton.1995}.}\BibitemShut {Stop}%
\bibitem [{\citenamefont {Siu}\ \emph {et~al.}(2012)\citenamefont {Siu},
  \citenamefont {Pluhackova},\ and\ \citenamefont {B{\"o}ckmann}}]{Siu.2012}%
  \BibitemOpen
  \bibfield  {author} {\bibinfo {author} {\bibfnamefont {S.~W.~I.}\
  \bibnamefont {Siu}}, \bibinfo {author} {\bibfnamefont {K.}~\bibnamefont
  {Pluhackova}}, \ and\ \bibinfo {author} {\bibfnamefont {R.~A.}\ \bibnamefont
  {B{\"o}ckmann}},\ }\href@noop {} {\bibfield  {journal} {\bibinfo  {journal}
  {J. Chem. Theory Comput.}\ }\textbf {\bibinfo {volume} {8}},\ \bibinfo
  {pages} {1459} (\bibinfo {year} {2012})}\BibitemShut {NoStop}%
\bibitem [{\citenamefont {Fetters}\ \emph
  {et~al.}(1999{\natexlab{a}})\citenamefont {Fetters}, \citenamefont {Lohse},
  \citenamefont {Milner},\ and\ \citenamefont {Graessley}}]{Fetters.1999}%
  \BibitemOpen
  \bibfield  {author} {\bibinfo {author} {\bibfnamefont {L.~J.}\ \bibnamefont
  {Fetters}}, \bibinfo {author} {\bibfnamefont {D.~J.}\ \bibnamefont {Lohse}},
  \bibinfo {author} {\bibfnamefont {S.~T.}\ \bibnamefont {Milner}}, \ and\
  \bibinfo {author} {\bibfnamefont {W.~W.}\ \bibnamefont {Graessley}},\
  }\href@noop {} {\bibfield  {journal} {\bibinfo  {journal} {Macromolecules}\
  }\textbf {\bibinfo {volume} {32}},\ \bibinfo {pages} {6847} (\bibinfo {year}
  {1999}{\natexlab{a}})}\BibitemShut {NoStop}%
\bibitem [{\citenamefont {Fetters}\ \emph
  {et~al.}(1999{\natexlab{b}})\citenamefont {Fetters}, \citenamefont {Lohse},\
  and\ \citenamefont {Graessley}}]{Fetters.1999.2}%
  \BibitemOpen
  \bibfield  {author} {\bibinfo {author} {\bibfnamefont {L.~J.}\ \bibnamefont
  {Fetters}}, \bibinfo {author} {\bibfnamefont {D.~J.}\ \bibnamefont {Lohse}},
  \ and\ \bibinfo {author} {\bibfnamefont {W.~W.}\ \bibnamefont {Graessley}},\
  }\href@noop {} {\bibfield  {journal} {\bibinfo  {journal} {J. Polym. Sci.
  Pol. Phys.}\ }\textbf {\bibinfo {volume} {37}},\ \bibinfo {pages} {1023}
  (\bibinfo {year} {1999}{\natexlab{b}})}\BibitemShut {NoStop}%
\bibitem [{\citenamefont {Jorgensen}\ \emph {et~al.}(1996)\citenamefont
  {Jorgensen}, \citenamefont {Maxwell},\ and\ \citenamefont
  {Tirado-Rives}}]{Jorgensen.1996}%
  \BibitemOpen
  \bibfield  {author} {\bibinfo {author} {\bibfnamefont {W.~L.}\ \bibnamefont
  {Jorgensen}}, \bibinfo {author} {\bibfnamefont {D.~S.}\ \bibnamefont
  {Maxwell}}, \ and\ \bibinfo {author} {\bibfnamefont {J.}~\bibnamefont
  {Tirado-Rives}},\ }\href@noop {} {\bibfield  {journal} {\bibinfo  {journal}
  {J. Am. Chem. Soc.}\ }\textbf {\bibinfo {volume} {118}},\ \bibinfo {pages}
  {11225} (\bibinfo {year} {1996})}\BibitemShut {NoStop}%
\bibitem [{\citenamefont {Sirk}\ \emph {et~al.}(2012)\citenamefont {Sirk},
  \citenamefont {Slizoberg}, \citenamefont {Brennan}, \citenamefont {Lisal},\
  and\ \citenamefont {Andzelm}}]{Sirk.2012}%
  \BibitemOpen
  \bibfield  {author} {\bibinfo {author} {\bibfnamefont {T.~W.}\ \bibnamefont
  {Sirk}}, \bibinfo {author} {\bibfnamefont {Y.~R.}\ \bibnamefont {Slizoberg}},
  \bibinfo {author} {\bibfnamefont {J.~K.}\ \bibnamefont {Brennan}}, \bibinfo
  {author} {\bibfnamefont {M.}~\bibnamefont {Lisal}}, \ and\ \bibinfo {author}
  {\bibfnamefont {J.~W.}\ \bibnamefont {Andzelm}},\ }\href {\doibase
  http://dx.doi.org/10.1063/1.3698476} {\bibfield  {journal} {\bibinfo
  {journal} {J. Chem. Phys.}\ }\textbf {\bibinfo {volume} {136}},\ \bibinfo
  {eid} {134903} (\bibinfo {year} {2012})}\BibitemShut {NoStop}%
\bibitem [{\citenamefont {Depa}\ \emph {et~al.}(2011)\citenamefont {Depa},
  \citenamefont {Chen},\ and\ \citenamefont {Maranas}}]{Maranas.2011}%
  \BibitemOpen
  \bibfield  {author} {\bibinfo {author} {\bibfnamefont {P.}~\bibnamefont
  {Depa}}, \bibinfo {author} {\bibfnamefont {C.}~\bibnamefont {Chen}}, \ and\
  \bibinfo {author} {\bibfnamefont {J.~K.}\ \bibnamefont {Maranas}},\ }\href
  {\doibase http://dx.doi.org/10.1063/1.3513365} {\bibfield  {journal}
  {\bibinfo  {journal} {J. Chem. Phys.}\ }\textbf {\bibinfo {volume} {134}},\
  \bibinfo {eid} {014903} (\bibinfo {year} {2011})}\BibitemShut {NoStop}%
\bibitem [{\citenamefont {Depa}\ and\ \citenamefont
  {Maranas}(2005)}]{Maranas.2005}%
  \BibitemOpen
  \bibfield  {author} {\bibinfo {author} {\bibfnamefont {P.~K.}\ \bibnamefont
  {Depa}}\ and\ \bibinfo {author} {\bibfnamefont {J.~K.}\ \bibnamefont
  {Maranas}},\ }\href {\doibase http://dx.doi.org/10.1063/1.1997150} {\bibfield
   {journal} {\bibinfo  {journal} {J. Chem. Phys.}\ }\textbf {\bibinfo {volume}
  {123}},\ \bibinfo {eid} {094901} (\bibinfo {year} {2005})}\BibitemShut
  {NoStop}%
\bibitem [{\citenamefont {Lyubimov}\ and\ \citenamefont
  {Guenza}(2013)}]{Lyubimov.2013}%
  \BibitemOpen
  \bibfield  {author} {\bibinfo {author} {\bibfnamefont {I.~Y.}\ \bibnamefont
  {Lyubimov}}\ and\ \bibinfo {author} {\bibfnamefont {M.~G.}\ \bibnamefont
  {Guenza}},\ }\href {\doibase http://dx.doi.org/10.1063/1.4792367} {\bibfield
  {journal} {\bibinfo  {journal} {J. Chem. Phys.}\ }\textbf {\bibinfo {volume}
  {138}},\ \bibinfo {eid} {12A546} (\bibinfo {year} {2013})}\BibitemShut
  {NoStop}%
\bibitem [{\citenamefont {Lyubimov}\ \emph {et~al.}(2010)\citenamefont
  {Lyubimov}, \citenamefont {McCarty}, \citenamefont {Clark},\ and\
  \citenamefont {Guenza}}]{Lyubimov.2010}%
  \BibitemOpen
  \bibfield  {author} {\bibinfo {author} {\bibfnamefont {I.~Y.}\ \bibnamefont
  {Lyubimov}}, \bibinfo {author} {\bibfnamefont {J.}~\bibnamefont {McCarty}},
  \bibinfo {author} {\bibfnamefont {A.}~\bibnamefont {Clark}}, \ and\ \bibinfo
  {author} {\bibfnamefont {M.~G.}\ \bibnamefont {Guenza}},\ }\href {\doibase
  http://dx.doi.org/10.1063/1.3450301} {\bibfield  {journal} {\bibinfo
  {journal} {J. Chem. Phys.}\ }\textbf {\bibinfo {volume} {132}},\ \bibinfo
  {eid} {224903} (\bibinfo {year} {2010})}\BibitemShut {NoStop}%
\bibitem [{\citenamefont {Harmandaris}(2014)}]{Harmandaris.2014}%
  \BibitemOpen
  \bibfield  {author} {\bibinfo {author} {\bibfnamefont {V.}~\bibnamefont
  {Harmandaris}},\ }\href {http://dx.doi.org/10.1007/s13367-014-0003-7}
  {\bibfield  {journal} {\bibinfo  {journal} {Korea-Aust. Rheol. J.}\ }\textbf
  {\bibinfo {volume} {26}},\ \bibinfo {pages} {15} (\bibinfo {year}
  {2014})}\BibitemShut {NoStop}%
\bibitem [{\citenamefont {Harmandaris}\ and\ \citenamefont
  {Kremer}(2009)}]{Harmandaris.2009}%
  \BibitemOpen
  \bibfield  {author} {\bibinfo {author} {\bibfnamefont {V.~A.}\ \bibnamefont
  {Harmandaris}}\ and\ \bibinfo {author} {\bibfnamefont {K.}~\bibnamefont
  {Kremer}},\ }\href@noop {} {\bibfield  {journal} {\bibinfo  {journal} {Soft
  Matter}\ }\textbf {\bibinfo {volume} {5}},\ \bibinfo {pages} {3920} (\bibinfo
  {year} {2009})}\BibitemShut {NoStop}%
\bibitem [{\citenamefont {Fritz}\ \emph {et~al.}(2011)\citenamefont {Fritz},
  \citenamefont {Koschke}, \citenamefont {Harmandaris}, \citenamefont {van~der
  Vegt},\ and\ \citenamefont {Kremer}}]{Fritz.2011}%
  \BibitemOpen
  \bibfield  {author} {\bibinfo {author} {\bibfnamefont {D.}~\bibnamefont
  {Fritz}}, \bibinfo {author} {\bibfnamefont {K.}~\bibnamefont {Koschke}},
  \bibinfo {author} {\bibfnamefont {V.~A.}\ \bibnamefont {Harmandaris}},
  \bibinfo {author} {\bibfnamefont {N.~F.~A.}\ \bibnamefont {van~der Vegt}}, \
  and\ \bibinfo {author} {\bibfnamefont {K.}~\bibnamefont {Kremer}},\
  }\href@noop {} {\bibfield  {journal} {\bibinfo  {journal} {Phys. Chem. Chem.
  Phys.}\ }\textbf {\bibinfo {volume} {13}},\ \bibinfo {pages} {10412}
  (\bibinfo {year} {2011})}\BibitemShut {NoStop}%
\bibitem [{\citenamefont {Hess}\ \emph {et~al.}(2006)\citenamefont {Hess},
  \citenamefont {Leon}, \citenamefont {van~der Vegt},\ and\ \citenamefont
  {Kremer}}]{Hess.2006}%
  \BibitemOpen
  \bibfield  {author} {\bibinfo {author} {\bibfnamefont {B.}~\bibnamefont
  {Hess}}, \bibinfo {author} {\bibfnamefont {S.}~\bibnamefont {Leon}}, \bibinfo
  {author} {\bibfnamefont {N.}~\bibnamefont {van~der Vegt}}, \ and\ \bibinfo
  {author} {\bibfnamefont {K.}~\bibnamefont {Kremer}},\ }\href {\doibase
  10.1039/B602076C} {\bibfield  {journal} {\bibinfo  {journal} {Soft Matter}\
  }\textbf {\bibinfo {volume} {2}},\ \bibinfo {pages} {409} (\bibinfo {year}
  {2006})}\BibitemShut {NoStop}%
\bibitem [{\citenamefont {de~Gennes}(1980)}]{deGennes.1980}%
  \BibitemOpen
  \bibfield  {author} {\bibinfo {author} {\bibfnamefont {P.~G.}\ \bibnamefont
  {de~Gennes}},\ }\href@noop {} {\bibfield  {journal} {\bibinfo  {journal} {J.
  Chem. Phys.}\ }\textbf {\bibinfo {volume} {72}},\ \bibinfo {pages} {4756}
  (\bibinfo {year} {1980})}\BibitemShut {NoStop}%
\bibitem [{\citenamefont {Vega}\ \emph {et~al.}(2004)\citenamefont {Vega},
  \citenamefont {Rastogi}, \citenamefont {Peters},\ and\ \citenamefont
  {Meijer}}]{Vega.2004}%
  \BibitemOpen
  \bibfield  {author} {\bibinfo {author} {\bibfnamefont {J.~F.}\ \bibnamefont
  {Vega}}, \bibinfo {author} {\bibfnamefont {S.}~\bibnamefont {Rastogi}},
  \bibinfo {author} {\bibfnamefont {G.~W.~M.}\ \bibnamefont {Peters}}, \ and\
  \bibinfo {author} {\bibfnamefont {H.~E.~H.}\ \bibnamefont {Meijer}},\ }\href
  {\doibase http://dx.doi.org/10.1122/1.1718367} {\bibfield  {journal}
  {\bibinfo  {journal} {J. Rheol.}\ }\textbf {\bibinfo {volume} {48}},\
  \bibinfo {pages} {663} (\bibinfo {year} {2004})}\BibitemShut {NoStop}%
\bibitem [{\citenamefont {Raju}\ \emph {et~al.}(1979)\citenamefont {Raju},
  \citenamefont {Smith}, \citenamefont {Marin}, \citenamefont {Knox},\ and\
  \citenamefont {Graessley}}]{Raju.1979}%
  \BibitemOpen
  \bibfield  {author} {\bibinfo {author} {\bibfnamefont {V.~R.}\ \bibnamefont
  {Raju}}, \bibinfo {author} {\bibfnamefont {G.~G.}\ \bibnamefont {Smith}},
  \bibinfo {author} {\bibfnamefont {G.}~\bibnamefont {Marin}}, \bibinfo
  {author} {\bibfnamefont {J.~R.}\ \bibnamefont {Knox}}, \ and\ \bibinfo
  {author} {\bibfnamefont {W.~W.}\ \bibnamefont {Graessley}},\ }\href
  {http://dx.doi.org/10.1002/pol.1979.180170704} {\bibfield  {journal}
  {\bibinfo  {journal} {J. Polym. Sci. Pol. Phys.}\ }\textbf {\bibinfo {volume}
  {17}},\ \bibinfo {pages} {1183} (\bibinfo {year} {1979})}\BibitemShut
  {NoStop}%
\bibitem [{\citenamefont {Moore}\ \emph {et~al.}(2014)\citenamefont {Moore},
  \citenamefont {Iacovella},\ and\ \citenamefont {McCabe}}]{Moore.2014}%
  \BibitemOpen
  \bibfield  {author} {\bibinfo {author} {\bibfnamefont {T.~C.}\ \bibnamefont
  {Moore}}, \bibinfo {author} {\bibfnamefont {C.~R.}\ \bibnamefont
  {Iacovella}}, \ and\ \bibinfo {author} {\bibfnamefont {C.}~\bibnamefont
  {McCabe}},\ }\href {\doibase http://dx.doi.org/10.1063/1.4880555} {\bibfield
  {journal} {\bibinfo  {journal} {J. Chem. Phys.}\ }\textbf {\bibinfo {volume}
  {140}},\ \bibinfo {eid} {224104} (\bibinfo {year} {2014})}\BibitemShut
  {NoStop}%
\bibitem [{\citenamefont {Carbone}\ \emph {et~al.}(2008)\citenamefont
  {Carbone}, \citenamefont {Varzaneh}, \citenamefont {Chen},\ and\
  \citenamefont {Müller-Plathe}}]{Carbone.2008}%
  \BibitemOpen
  \bibfield  {author} {\bibinfo {author} {\bibfnamefont {P.}~\bibnamefont
  {Carbone}}, \bibinfo {author} {\bibfnamefont {H.~A.~K.}\ \bibnamefont
  {Varzaneh}}, \bibinfo {author} {\bibfnamefont {X.}~\bibnamefont {Chen}}, \
  and\ \bibinfo {author} {\bibfnamefont {F.}~\bibnamefont {Müller-Plathe}},\
  }\href {\doibase http://dx.doi.org/10.1063/1.2829409} {\bibfield  {journal}
  {\bibinfo  {journal} {J. Chem. Phys.}\ }\textbf {\bibinfo {volume} {128}},\
  \bibinfo {eid} {064904} (\bibinfo {year} {2008})}\BibitemShut {NoStop}%
\bibitem [{\citenamefont {Hoy}\ and\ \citenamefont
  {Karayiannis}(2013)}]{Hoy.2013}%
  \BibitemOpen
  \bibfield  {author} {\bibinfo {author} {\bibfnamefont {R.~S.}\ \bibnamefont
  {Hoy}}\ and\ \bibinfo {author} {\bibfnamefont {N.~C.}\ \bibnamefont
  {Karayiannis}},\ }\href {\doibase 10.1103/PhysRevE.88.012601} {\bibfield
  {journal} {\bibinfo  {journal} {Phys. Rev. E}\ }\textbf {\bibinfo {volume}
  {88}},\ \bibinfo {pages} {012601} (\bibinfo {year} {2013})}\BibitemShut
  {NoStop}%
\bibitem [{\citenamefont {Tuckerman}\ \emph {et~al.}(1992)\citenamefont
  {Tuckerman}, \citenamefont {Berne},\ and\ \citenamefont
  {Martyna}}]{Tuckerman.1992}%
  \BibitemOpen
  \bibfield  {author} {\bibinfo {author} {\bibfnamefont {M.}~\bibnamefont
  {Tuckerman}}, \bibinfo {author} {\bibfnamefont {B.~J.}\ \bibnamefont
  {Berne}}, \ and\ \bibinfo {author} {\bibfnamefont {G.~J.}\ \bibnamefont
  {Martyna}},\ }\href@noop {} {\bibfield  {journal} {\bibinfo  {journal} {J.
  Chem. Phys.}\ }\textbf {\bibinfo {volume} {97}},\ \bibinfo {pages} {1990}
  (\bibinfo {year} {1992})}\BibitemShut {NoStop}%
\bibitem [{\citenamefont {Isele-Holder}\ \emph {et~al.}(2012)\citenamefont
  {Isele-Holder}, \citenamefont {Mitchell},\ and\ \citenamefont
  {Ismail}}]{Isele-Holder.2012}%
  \BibitemOpen
  \bibfield  {author} {\bibinfo {author} {\bibfnamefont {R.~E.}\ \bibnamefont
  {Isele-Holder}}, \bibinfo {author} {\bibfnamefont {W.}~\bibnamefont
  {Mitchell}}, \ and\ \bibinfo {author} {\bibfnamefont {A.~E.}\ \bibnamefont
  {Ismail}},\ }\href {\doibase http://dx.doi.org/10.1063/1.4764089} {\bibfield
  {journal} {\bibinfo  {journal} {J. Chem. Phys.}\ }\textbf {\bibinfo {volume}
  {137}},\ \bibinfo {eid} {174107} (\bibinfo {year} {2012})}\BibitemShut
  {NoStop}%
\bibitem [{\citenamefont {Plimpton}(1995)}]{Plimpton.1995}%
  \BibitemOpen
  \bibfield  {author} {\bibinfo {author} {\bibfnamefont {S.}~\bibnamefont
  {Plimpton}},\ }\href@noop {} {\bibfield  {journal} {\bibinfo  {journal} {J.
  Comput. Phys.}\ }\textbf {\bibinfo {volume} {117}},\ \bibinfo {pages} {1 }
  (\bibinfo {year} {1995})}\BibitemShut {NoStop}%
\end{thebibliography}

\end{document}